\documentstyle[aps,prl,epsf]{revtex}
\begin{document}
\newcommand{\identity}{\:\mbox{\sf 1} \hspace{-0.37em} \mbox{\sf 1}\,}
\newcommand{\ket}[1]{| #1 \rangle}
\newcommand{\bra}[1]{\langle #1 |}
\newcommand{\proj}[1]{| #1\rangle\!\langle #1|}


\author{John A. Smolin}

\title{A four-party unlockable bound-entangled state}

\address{IBM T.J. Watson Research Center, Yorktown Heights, NY 10598,
smolin@watson.ibm.com
\\}

\date{January 1, 1900}

\maketitle
\begin{abstract}
I present a four-party unlockable bound-entangled state, that is, a
four-party quantum state which cannot be written in a separable form
and from which no pure entanglement can be distilled by local quantum
operations and classical communication among the parties, and yet when
any two of the parties come together in the same laboratory they can
perform a measurement which enables the other two parties to create a
pure maximally entangled state between them without coming together.
This unlocking ability can be viewed in two ways, as either a
determination of which Bell state is shared in the mixture, or as a
kind of quantum teleportation with cancellation of Pauli operators.
\end{abstract}

%

The study of entanglement, the so-called ``spooky action at a
distance'' of quantum particles whose joint states cannot be written
in a product form \cite{epr}, has been at the heart of quantum
information theory, and seems to be crucial to an understanding of
quantum computation, quantum cryptography and perhaps quantum
mechanics itself.  It has been shown that in the case of mixed
entangled states, it is often possible to distill some nearly pure
entanglement using only local quantum operations and classical
communications among the parties sharing the state
\cite{purification,bdsw}.  Recently, a new type of entangled mixed
state was discovered \cite{horodecki,horodeckis} which has the
property that, though definitely entangled, is not distillable.  Such
states are known as {\em bound entangled} states.

The usual technique used in proofs about bound entanglement to show a
state is entangled is to observe that there are not enough product
states in its span for it to be decomposed in separable form:
\begin{equation} 
\sum_i \alpha_i \proj{\psi^1_i} \otimes \proj{\psi^2_i}\otimes \proj{\psi^3_i}
\otimes \ldots \proj{\psi^N_i}
\label{separableform}
\end{equation}
where there are tensor Hilbert spaces 1 through $N$ and the $\alpha_i$'s 
are real numbers summing to 1.

This note will show a state is entangled by a different method, by
showing that when two parties of a four-party state come together,
they can by local quantum operations and classical communication
enable the other two parties to have some pure entanglement (for a
discussion of multiparty entanglement purification protocols see
\cite{mppvk,MS}).  It will further be shown that this entanglement is
not available without the coming together of two of the parties, thus
the state is bound entangled.  These results may have applications to 
quantum cryptography (if two parties manage to share a pure maximally 
entangled state they can also share secure key bits) and quantum 
secret sharing \cite{qne,secretsharing} in a multi-party setting.

The unlockable state is:
\begin{eqnarray}
\nonumber \rho=\frac{1}{4} \left(
\ket{\Phi^+}^{AB}\bra{\Phi^+}\otimes\ket{\Phi^+}^{CD}\bra{\Phi^+}\ +
\ \ket{\Phi^-}^{AB}\bra{\Phi^-}\otimes\ket{\Phi^-}^{CD}\bra{\Phi^-}\ +\right.\\
\left.\ket{\Psi^+}^{AB}\bra{\Psi^+}\otimes\ket{\Psi^+}^{CD}\bra{\Psi^+}\ +
\ \ket{\Psi^-}^{AB}\bra{\Psi^-}\otimes\ket{\Psi^-}^{CD}\bra{\Psi^-} \right)
\label{thestate}
\end{eqnarray}
where we use the usual notation for the maximally entangled states
of two qubits (the Bell states):
\begin{equation}
\ket{\Psi^\pm}=\frac{1}{\sqrt{2}}\left(\ket{\!\uparrow\downarrow}\pm\ket{\!\uparrow\downarrow}\right),
\ \ket{\Phi^\pm}=\frac{1}{\sqrt{2}}\left(\ket{\!\uparrow\uparrow}\pm\ket{\!\downarrow\downarrow}\right)
\end{equation}
In other words, $A$ and $B$ share one of the four Bell states, but don't
know which one, and $C$ and $D$ share the same Bell state, also
not knowing which it is.

If $C$ and $D$ come together into the same laboratory and do
the nonlocal Bell measurement on their systems, they can determine
reliably which Bell state they had since the four Bell states are
orthogonal.  They can then send this classical information to
$A$ and $B$ who will then know which Bell state they have
and can convert it into the standard state $\ket{\Psi^-}$
unitarily and locally using the following relations, up to an
unimportant overall phase:
\begin{equation}
\begin{array}{l}
\ket{\Psi^-}\propto \identity_2 \otimes \sigma_0 \ket{\Psi^-} \propto
\sigma_0 \otimes \identity_2 \ket{\Psi^-}\\
\ket{\Psi^-}\propto \identity_2 \otimes \sigma_1 \ket{\Psi^+} \propto
\sigma_1 \otimes \identity_2 \ket{\Psi^+}\\
\ket{\Psi^-}\propto \identity_2 \otimes \sigma_2 \ket{\Phi^+} \propto
\sigma_2 \otimes \identity_2 \ket{\Phi^+}\\
\ket{\Psi^-}\propto \identity_2 \otimes \sigma_3 \ket{\Phi^-} \propto
\sigma_3 \otimes \identity_2 \ket{\Phi^-}
\end{array}
\label{propto}
\end{equation}
where the $\sigma$'s are the members of the set of rotation matrices
$\sigma=\{\identity_2,{1\ \ 0 \choose 0\ -1},{0\ -1 \choose 1\ \ 0}, 
{0\ 1 \choose 1\ 0}\}$. 
The Rotations $\sigma$ are simply the identity and the three Pauli spin
operators, leaving aside imaginary parts which contribute only to the
overall phase.  The single singlet obtained between $A$ and $B$
by this procedure is all that can be distilled.  This is a simple
consequence of $A$ and $B$ each possessing only one qubit in
the original state $\rho$.

Because entanglement between $A$ and $B$ can be distilled from it, 
$\rho$ must be entangled. If it were not it could be written in the 
biseparable form
\begin{equation}
\rho=\sum_i \alpha_i 
\proj{\psi_i^A}\otimes\proj{\phi_i^{BCD}}\ .
\label{sep2}
\end{equation}
It was proven in \cite{bdsw} that if two parties are on opposite sides
of a separable cut, then local quantum operations and classical
communication will always leave them in a separable form, which
implies immediately that no pure entanglement can be distilled between
them.  So if $\rho$ is of the form (\ref{sep2}) there would be no way
to distill any entanglement between $A$ and any of the other parties,
including $B$, even if all three other parties $B$, $C$ and $D$ join
together.  Since it actually is possible to distill entanglement under
these conditions (having $B$ in the same laboratory with $C$ and $D$
can only help) $\rho$ must have been entangled all along.

On the other hand, if all four parties remain in separate labs the
state is not distillable.  The proof of this will be based on looking
at various cuts across which $\rho$ is separable, despite the fact
that it is an entangled state.  To demonstrate the nondistillability
of $\rho$ it will be sufficient to show that, despite being entangled,
$\rho$ is separable across the three bipartite cuts $AB:CD$, $AC:BD$
and $AD:BC$.  This will separate every party from every other party,
and every pair of parties from every pair, across at least one
separable boundary.  This requires that no entanglement can be
distilled between any two parties or any two pairs, leaving only the
possibility of distilling some three- or four-party entanglement.
This is ruled out by noting that any such entanglement would span a
separable bipartite cut.  For example, if there were some distilled
$A:BCD$ entanglement, it would still have to be separable across the
$AB:CD$ boundary, leaving only the possibility of some entanglement of
$A$ with $B$ and/or some entanglement of $C$ with $D$, each of which
has already been excluded.

The state $\rho$ is separable across the $AB:CD$ boundary as it is written in
separable form (\ref{thestate}).  One way to show the state is
separable across the $AC:BD$ cut is to rewrite the state with $B$ and
$C$ interchanged and consider the original $AB:CD$ cut.  After
interchanging indices it is easy to show that $\rho$ is invariant
under the interchange of $B$ and $C$ and is therefore separable across
the $AC:BD$ cut.  Writing out each vector in the mixture (leaving out the $1/2$ 
normalization for clarity):
\begin{equation}
\begin{array}{c}
\ket{\Phi^+}^{AB}\otimes\ket{\Phi^+}^{CD}=(\ket{00}+\ket{11})\otimes(\ket{00}+\ket{11})=
\ket{0000}+\ket{0011}+\ket{1100}+\ket{1111}\\
\ket{\Phi^-}^{AB}\otimes\ket{\Phi^-}^{CD}=(\ket{00}-\ket{11})\otimes(\ket{00}-\ket{11})=
\ket{0000}-\ket{0011}-\ket{1100}+\ket{1111}\\

\ket{\Psi^+}^{AB}\otimes\ket{\Psi^+}^{CD}=(\ket{01}+\ket{10})\otimes(\ket{01}+\ket{10})=
\ket{0101}+\ket{0110}+\ket{1001}+\ket{1010}\\
\ket{\Psi^-}^{AB}\otimes\ket{\Psi^-}^{CD}=(\ket{01}-\ket{10})\otimes(\ket{01}-\ket{10})=
\ket{0101}-\ket{0110}-\ket{1001}+\ket{1010}\\
\end{array}\ .
\label{big1}
\end{equation}  
Now, by interchanging the $B$ and $C$ index we have
\begin{equation}
\begin{array}{c}
\ket{\Phi^+}^{AC}\otimes\ket{\Phi^+}^{BD}=
\ket{0000}+\ket{0101}+\ket{1010}+\ket{1111}\\
\ket{\Phi^-}^{AC}\otimes\ket{\Phi^-}^{BD}=
\ket{0000}-\ket{0101}-\ket{1010}+\ket{1111}\\

\ket{\Psi^+}^{AC}\otimes\ket{\Psi^+}^{BD}=
\ket{0011}+\ket{0110}+\ket{1001}+\ket{1100}\\
\ket{\Psi^-}^{AC}\otimes\ket{\Psi^-}^{BD}=
\ket{0011}-\ket{0110}-\ket{1001}+\ket{1100}\\
\end{array}\ .
\label{big2}
\end{equation}  
First note that in both cases when the outer product is taken and the
projectors corresponding to these vectors are mixed together, all the
minus signs will vanish.  Terms with minus signs combined with each
other will have the sign cancel.  Negative terms combined with
positive terms will be cancelled since all the negative terms appear
elsewhere as positive terms.  So either the signs or the cross-terms
having them all cancel, and we can ignore sign hereafter.  It is then
simple to check that every term in (\ref{big1}) also appears in
(\ref{big2}), just in a different place.  When the projectors are
added up they will result in the same final density matrix.  The same
property will hold for the $AD:BC$ cut which is symmetric with the
$AC:BD$ case.  Thus, $\rho$ has been shown to be not distillable and
therefore its entanglement is bound.

If $\rho$ is separable across the $AC:BD$ cut, for instance, how is
it possible that $C$ and $D$ coming together can enable $A$ and 
$B$ to become entangled?  The answer is that when $C$ and $D$ join
together in the same laboratory, they have crossed the line of the
cut and can create obviously entanglement across it.  The surprising
thing is that this entanglement is not only shared by $C$ and $D$ 
but by $A$ and $B$.  It would not have been possible for $A$ and $B$ 
to become entangled without themselves getting together in the same
laboratory were $\rho$ entirely four-way separable (\ref{separableform}) 
to begin with, so the whole process depends on $\rho$'s having some 
four-way entanglement.  

The invariance under interchange of particles noted above also makes
it clear that $\rho$ has the property that if any two of the parties
come together they can perform the Bell measurement and pass classical
information to the other two parties giving them a distilled Bell
state.  Since it is not immediately obvious why this distillation
works when, for example, $B$ and $D$ get together, since they don't as
clearly share a Bell state containing information about which Bell
state the others share as when $A$ and $B$ or $C$ and $D$ get
together, it is instructive to look at an alternative explanation for
what is going on.

\begin{figure}
\epsfxsize=14cm
\epsfbox{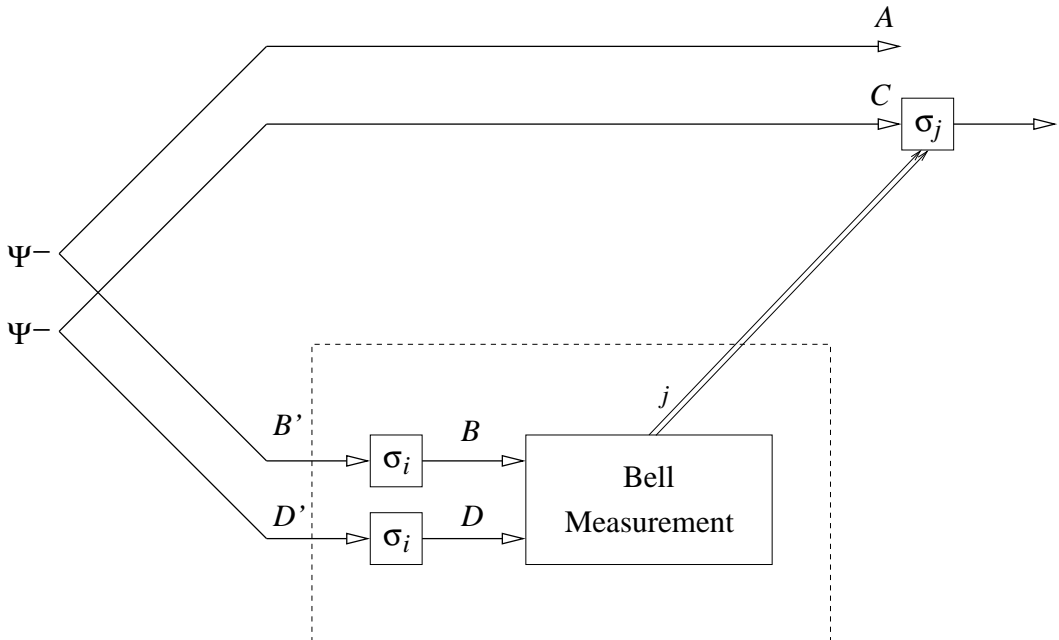}
\caption{$A$ and $B$ (and $C$ and $D$)
share a $\ket{\Psi^-}$ which has been turned into
one of the four possible Bell states by $\sigma_i$.  When the $\sigma_i$'s are
merged into the Bell measurement, we have teleportation from $B'$ to $C$.
}
\label{teleport}
\end{figure}

Since all the $\sigma_i$'s are, up to a phase, self-inverse, and since
Eq. (\ref{propto}) works whichever party applies the rotation, it must
be that the $\sigma_i$'s can be used in reverse, to create one of the other
Bell states out of a $\ket{\Psi^-}$.  This is illustrated in Figure
\ref{teleport}.  The Bell measurement is just a rotation to the Bell
basis (made up of a matrix whose rows are the Bell states) followed
by a measurement in the standard basis.  If we now think of the
$\sigma_i$'s as multiplying the rows of the Bell measurement on the
right rather than the original $\ket{\Psi^-}$'s on the left, we can
see that they cancel each other out, up to a phase, and the resulting
measurement inside the dashed box is the same as the original Bell
measurement.  We can then think of the whole procedure as $B$ and $D$
getting together to {\em teleport} \cite{teleportation} half of a
$\ket{\Psi^-}$ belonging to $A$ and $B'$ to $C$ using the
$\ket{\Psi^-}$ shared by $C$ and $D'$.  The measurement will result in
two bits of classical data $j$ which will be used at $C$ to complete
the teleportation by performing a $\sigma_j$ rotation in exactly the same
way as in Eq. (\ref{propto}).  Thus we may think of the whole process
as either two parties measuring which Bell state they have
(determining the unknown $\sigma_i$) or as their teleporting half of a 
$\ket{\Psi^-}$ they share with one party to the remaining party,
with an implicit cancellation of the $\sigma_i$'s.

The ``unlocking'' feature, that two parties can assist the other two
in getting some entanglement, is reminiscent of the unlocking of
hidden entanglement discussed by Cohen \cite{cohen} also known as the
entanglement of assistance \cite{assistance}.  The new feature here is
that the unlockable four-party state is bound entangled--the
entanglement is not available if none of the parties can perform joint
quantum operations.  The earlier examples explicitly allow one of
three parties, say $C$, to give the other two parties $A$ and $B$ some
classical information which they can use to obtain some pure
entanglement even though the joint state of $A$ and $B$ ignoring $C$
is separable, thus these are examples of three-party distillable
states.  These are two distinct types of unlocking: In one case $C$
can unlock the hidden entanglement shared by $A$ and $B$; in the other
the {\em ability} of $C$ and $D$ to unlock the entanglement of $A$ and $B$ is 
itself unlocked by their coming together.  

In \cite{dur1}, D\"ur, Cirac and Tarrach give a three-qubit state with
the property that it is $A:BC$ and $B:AC$ separable, but not separable
$AB:C$ (what they call a class 3 state).  These separability conditions
are sufficient to show using arguments similar to the above that their
state is not distillable when all the parties are isolated.  They
further point out that their state has negative partial transpose with
respect to C and that the state is therefore AB:C distillable because
any state in $2\otimes n$ with negative partial transpose is
distillable \cite{durbruss}.  Thus, they have provided the first example
of an unlockable bound-entangled state, though it will require many
copies of the state to perform the distillation, lacking the direct
distillability in one copy of $\rho$.

The type of unlocking exhibited by D\"ur, Cirac and Tarrach state differs 
subtly from that of $\rho$ presented here:  In their state, when $A$ and $B$ 
get together it is {\em they} who gain distillable entanglement with
$C$.  On the other hand, in the case of the four-party state, when 
two of the parties get together they gain nothing themselves,
merely the ability to give the other two parties distillable entanglement.  
If three of the four parties of $\rho$ get together, the situation 
will be that of the D\"ur, Cirac and Tarrach state.  This suggests 
the following categorization of states:  
\begin{itemize} 
\item Altruistic states: States where one party can help the others 
distill some entanglement, but gets none in return.  Examples of these 
are the states with hidden entanglement studied by Cohen 
\cite{cohen} and DiVincenzo et. al. \cite{assistance}.  In particular
the Greenberger, Horne and Zeilinger (GHZ) state \cite{GHZ} has this
property.
\item Unlockable bound-entangled states:  States that are bound
unless some parties come together, after which some entanglement
can be distilled between remaining separated parties.  These include
the D\"ur, Cirac and Tarrach state, as well as $\rho$.
\item Unlockable bound-altruistic states:  Bound-entangled states that when
some parties come together are reduced to altruistic states, $\rho$ 
being the first example.
\end{itemize}

Other states that have some multi-party entanglement, but are
separable across various cuts, have been studied in
\cite{dur1,bras-mor,upb1,upb2,dur2,dur3}.  The state $\rho$ has $A:B:C:D$
bound entanglement, when grouped $AB:C:D$ has distillable $C:D$ entanglement,
and is separable $AB:CD$, and similarly for all permutations of the parties.
A three-party state given in \cite{upb1} has $A:B:C$ bound entanglement
and is separable $A:BC$, $AB:C$ and $AC:B$.  

There are several obvious generalizations of unlockable states to
higher dimensions and more parties.  For example, a four-party state
of the same form as $\rho$ (Eq. (\ref{thestate})) but using the $n^2$
orthogonal maximally entangled states in $n\otimes n$ will have the
same properties: The unlocking measurement performed by $C$ and $D$ is
just a measurement in the basis of the maximally entangled states, the
separability across the $AB:CD$ cut is again by construction, and the
symmetry is easy to see using the teleportation argument with the
$\sigma_i$'s being the members of Heisenberg group in $n$ dimensions.

One could also look for states where if $n$ parties come together they
can cause the remaining $m$ to have entanglement, or some subset of
the remaining $m$, or where when some of the parties come together
they can cause the remaining parties to still have an unlockable bound
entangled state.  Some such states may be constructed by distributing
the parts of several copies of $\rho$ among several (more than four)
different parties.  Some surprises await, however: The tensor product
of two copies of $\rho$, one shared by the four parties $A$, $B$, $C$
and $D$ and another shared by $A$, $B$, $C$ and a fifth party $E$ can
be distilled into an EPR pair shared by $D$ and $E$, even though the
individual copies of $\rho$ are not distillable at all, providing an
example of superadditivity of distillable entanglement 
\cite{ashishandjohnandpeter}.  The many variations of such states and
their applications to the cryptographic ``web of trust'' are beyond
the scope of this letter, but will the the subject of future work.

A particular related question is whether there is an example of an
unlockable bound entangled state of rank lower than four.  It was
shown in \cite{norank2} that there exist no rank two bipartite bound
entangled states.  If a multi-partite bound entangled state were to
exist, it would have to be that when enough parties join together the
remaining bipartite state is always either separable or distillable.
Since we now see that there do exist states that become distillable
as parties join up, the search for a lower rank multi-party bound
entangled state may prove fruitful.

The author would like to thank Charles Bennett, Ignacio Cirac, David
DiVincenzo, Wolfgang D\"ur, Oliver Cohen, Barbara Terhal, 
and Ashish Thapliyal for helpful discussions, and the Army Research 
Office for support under contract number DAAG55-98-C-0041.
 
\begin{references}
\bibitem{epr}A. Einstein, B. Podolsky, N. Rosen, Phys. Rev. {\bf 47},
(1935) 777.
\bibitem{purification} C.H. Bennett, G. Brassard, S. Popescu,
B. Schumacher, J.A. Smolin, and W.K. Wootters,
Phys. Rev. Lett. {\bf 76}, 722 (1996).
\bibitem{bdsw}C.H. Bennett, D.P. DiVincenzo, J.A. Smolin, and W.K. Wootters,
Phys. Rev. A. {\bf 54}, 3824 (1996), LANL preprint quant-ph/9604024.
\bibitem{horodecki} P. Horodecki, Phys. Lett. A {\bf 232}, 333 (1997), 
LANL preprint quant-ph/9703004.
\bibitem{horodeckis} P. Horodecki, M. Horodecki, and R. Horodecki, Phys. Rev.
Lett. {\bf 80}, 5239 (1998), LANL preprint quant-ph/9801069. 
\bibitem{mppvk} M. Murao, M.B. Plenio, S. Popescu, V. Vedral, P.L. Knight 
Phys. Rev. A {\bf 57},  4075 (1998), LANL preprint quant-ph/9712045.
\bibitem{MS} E. Maneva and J.A. Smolin, ``Improved two-party and multi-party 
purification protocols,'' to appear in the AMS Contemporary Math Series 
volume entitled "Quantum Computation \& Quantum Information Science,"
LANL preprint quant-ph/0003099.
\bibitem{qne} C.H. Bennett, D.P. DiVincenzo, C.A. Fuchs, T. Mor,
E. Rains, P.W. Shor, J.A. Smolin, and W.K. Wootters, Phys. 
Rev. A {\bf 59}, 1070 (1999), LANL preprint quant-ph/9804053.
\bibitem{secretsharing} R. Cleve, D. Gottesman, and H.-K. Lo,
Phys. Rev. Lett. {\bf 83}, 648 (1999), LANL preprint quant-ph/9901025.
\bibitem{teleportation} C.H. Bennett, G. Brassard, C. Cr\'epeau, R. Jozsa,
A. Peres, and W.K. Wootters, Phys. Rev. Lett. {\bf 70}, 1895 (1993).
\bibitem{cohen} O. Cohen, Phys. Rev. Lett. {\bf 80}, 2493 (1998).
\bibitem{assistance} D.P. DiVincenzo, C.A. Fuchs,
H. Mabuchi, J.A. Smolin, A.V. Thapliyal, and A. Uhlmann,
Proc. 1st NASA Intl. Conf. on Quantum Computing and Quantum Communication,
Vol. 1509 of Lecture Notes in Computer Science (Springer), (1998),
LANL preprint quant-ph/9803033.
\bibitem{dur1} W. D\"ur, J.I. Cirac, and R. Tarrach, Phys. Rev. Lett. 
{\bf 83}, 3562 (1999), LANL preprint quant-ph/9903018.
\bibitem{durbruss} W. D\"ur, J.I. Cirac, M. Lewenstein, D. Bru\ss, 
LANL preprint quant-ph/9910022.
\bibitem{GHZ} D.M. Greenberger, M. Horne, A. Zeilinger, 
Am. J. Phys. {\bf 58}, 1131 (1990).
\bibitem{bras-mor} G. Brassard and T. Mor, ``Multi-particle entanglement via
2-particle entanglement,''
1st NASA Intl. Conf. on Quantum Computing and Quantum Communication,
Vol. 1509 of Lecture Notes in Computer Science (Springer), (1998).
\bibitem{upb1} C.H. Bennett, D.P. DiVincenzo, T. Mor, P.W. Shor,
J.A. Smolin and B.M. Terhal,  Phys. Rev. Lett. {\bf 82}, 5385 (1999),  
LANL preprint quant-ph/9808030.
\bibitem{upb2} D.P. DiVincenzo, T. Mor, P.W. Shor,
J.A. Smolin and B.M. Terhal, LANL preprint quant-ph/9908070.
\bibitem{dur2} W. D\"ur, J.I. Cirac, LANL preprint quant-ph/9911044. 
\bibitem{dur3} W. D\"ur, J.I. Cirac, LANL preprint quant-ph/0002028.
\bibitem{ashishandjohnandpeter} P.W. Shor, J.A. Smolin and A.V. Thapliyal, 
in preparation.
\bibitem{norank2} P. Horodecki, J.A. Smolin, B.M. Terhal and A.V. Thapliyal,
LANL preprint quant-ph/9910122.



\end{references}
\end{document}